\documentclass[aps,prd,twocolumn,groupedaddress,nofootinbib,showkeys,showpacs,sort&compress]{revtex4}

\usepackage{graphicx}

\newcommand{\escalpr}{\!\cdot\!}

\begin{document}

\title{Survey of $J=0,1$ mesons in a Bethe-Salpeter approach}

\author{A. Krassnigg}
\email[]{andreas.krassnigg@uni-graz.at}
\affiliation{Institut f\"ur Physik, Karl-Franzens-Universit\"at Graz, A-8010 Graz, Austria}

\date{\today}

\begin{abstract}
The Bethe-Salpeter equation is used to comprehensively study mesons with $J=0,1$ and equal-mass constituents for quark masses from the 
chiral limit to the $b$-quark mass. The survey contains masses of the ground states in all corresponding $J^{PC}$ channels
including those with ``exotic'' quantum numbers. The emphasis is put on each particular state's sensitivity 
to the low- and intermediate-momentum, i.e., long-range part of the strong interaction.
\end{abstract}


\maketitle

\section{Introduction\label{intro}}
Mesons offer a prime target for studies of various approaches to quantum chromodynamics (QCD), which
is widely accepted as the quantum field theory of the strong interaction. While in terms of the number of
constitutents their appearance is simple at first glance, mesons provide a broad range of phenomena and
challenges to both theory and experiment. On the theoretical side the key challenge is to understand 
mesons (and hadrons in general) as bound states of QCD's elementary degrees of freedom, quarks and gluons. 
The various approaches used to provide direct or indirect insight regarding this problem are (relativistic) quark 
models 
(e.\,g.~\cite{Godfrey:1985xj,Barnes:2005pb,Krassnigg:2003gh,Krassnigg:2004sp,Biernat:2007dn,Biernat:2009my,Nefediev:2006bm,Lakhina:2006vg,Lakhina:2006fy}
and references therein), reductions of the Bethe-Salpeter equation 
(e.\,g.~\cite{Koll:2000ke,Ricken:2003ua,Lucha:2005nz,Li:2007td} and references therein), 
lattice QCD (e.\,g.~\cite{McNeile:2006nv,Burch:2006dg,Wada:2007cp,Dudek:2007wv,Gattringer:2008be,Gregory:2008mn} 
and references therein), effective field theories (e.\,g.~\cite{Brambilla:2004wf} and references therein),
and the Dyson-Schwinger approach used herein.
On the experimental side, present-day challenges can be exemplified by the recent measurement
of the pseudoscalar ground-state mass in the bottomonium system \cite{Aubert:2008vj}.

In the present work, mesons are studied by means of QCDs Dyson-Schwinger-equations (DSEs);
for recent reviews, see \cite{Fischer:2006ub,Roberts:2007jh}. The DSEs are an infinite set of coupled 
and in general nonlinear integral equations for the Green functions of a quantum field theory, which 
makes the approach fully nonperturbative. It can therefore be used to study prominent
nonperturbative phenomena of QCD, namely dynamical chiral symmetry
breaking and confinement, as well as bound states in a single framework. The latter are studied in this approach
with the help of covariant equations. In particular, the Bethe-Salpeter equation (BSE) is used to 
describe a meson as a quark-antiquark system in QCD \cite{Smith:1969az}. Note that the analogous approach to 
baryons as systems of three spin-$1/2$ quarks is considerably more involved; a first realization of this
problem has been achieved only recently \cite{Eichmann:2009zx}. In the past, intermediate steps have been
taken to allow for the same level of sophistication as in corresponding meson studies (for recent advances, see
\cite{Eichmann:2007nn,Eichmann:2008ef,Nicmorus:2008vb,Nicmorus:2008eh,Eichmann:2008kk} and references therein). 

In principle one would
aim at a complete, self-consistent solution of all equations, which is equivalent to a solution of the 
underlying theory. While this spirit can be held up in investigations of certain aspects of the theory
(see, e.\,g.~\cite{Alkofer:2008tt,Fischer:2008uz} and references therein), numerical studies of hadronic 
observables require a truncation of the infinite tower of equations. In practice this means the choice of 
a subset of equations which are solved self-consistently by neglecting or making sophisticated Ans\"atze 
for the Green functions whose DSEs are not solved explicitly.

A popular truncation for meson studies in the DSE approach, which is also used in this work, is the rainbow-ladder (RL) truncation 
for reasons of simplicity and for satisfying among others the axial-vector Ward-Takahashi identity (AVWTI).
The AVWTI is a welcome restriction of the unknown quark-antiquark scattering kernel in the BSE
and its satisfaction guarantees the correct implementation of chiral symmetry and its dynamical 
breaking. A possible symmetry-preserving nonperturbative truncation scheme \cite{Munczek:1994zz,Bender:1996bb} contains
the RL truncation as the lowest order, various corrections to which can be systematically included in 
studies ``beyond RL'' \cite{Bhagwat:2004hn,Watson:2004kd,Watson:2004jq,Fischer:2005en,Matevosyan:2006bk,Fischer:2007ze,Fischer:2008wy,Fischer:2009jm}.
Another recent approach aims at the direct construction of the symmetry-preserving kernel of the BSE
from a general quark-gluon vertex \cite{Chang:2009zb}.
With chiral symmetry and its dynamical breaking correctly realized in this fashion and built into the 
calculation from the very beginning, one obtains a generalized Gell-Mann--Oakes--Renner relation valid
for all pseudoscalar mesons and all current-quark masses \cite{Maris:1997hd,Maris:1997tm,Holl:2004fr}.
In particular, the pion becomes massless in the chiral limit.

Meson studies in such a setup have been carried out over a number of years with various levels of
sophistication \cite{Munczek:1983dx,Jain:1993qh,Frank:1995uk,Burden:1997fq,Maris:1997tm,Maris:1999nt,Alkofer:2002bp}. 
Among these different variants the setup of Ref.~\cite{Maris:1999nt} has been successfully applied to the properties
of pseudoscalar and vector meson ground states, in particular electromagnetic form factors (see \cite{Bhagwat:2006pu} 
and references therein). Further applications include an exploratory study of hadronic meson decays \cite{Jarecke:2002xd},
calculations of diquark properties \cite{Maris:2002yu,Maris:2004bp} (since these correlations are of importance 
in baryon studies), and studies of radial meson excitations \cite{Holl:2004fr,Holl:2004un,Holl:2005vu,Krassnigg:2006ps,Bhagwat:2007rj}.

While early studies were conducted for light mesons, an extension to heavy-heavy mesons seemed natural 
\cite{Krassnigg:2004if,Maris:2005tt,Bhagwat:2006xi}, but required a change of method, since reaching the 
$b$-quark mass \cite{Maris:2006ea,Nguyen:2009if} is only possible with proper numerical treatment.

While many aspects of mesons with $J=0,1$ have been investigated separately, a comprehensive collection and discussion of 
the corresponding spectra is still missing. In the present work, as a first step, meson masses are presented for mesons
with equal mass constituents and all quantum numbers possible for $J=0,1$ for the cases of light, 
strange, charm, and bottom quark masses. The dependence of the masses on the parameters used in the interaction
are explored throughout, which appears to be an important issue for any such calculation.

The paper is organized as follows: Sec.~\ref{formalism} lists the necessary ingredients for the calculation, 
the results are presented and discussed in Sec.~\ref{results}, conclusions are offered in Sec.~\ref{conclusions}. 
The necessary details on the structure of the mesons' Bethe-Salpeter amplitudes for all quantum numbers considered 
here are collected in an appendix. All calculations have been performed in Euclidean momentum space.

\section{Gap equation, BSE, and interaction}\label{formalism}
In RL truncation one studies a meson with total $q\bar{q}$ momentum $P$ and relative $q\bar{q}$ 
momentum $k$ by consistently solving two equations: the homogeneous, ladder-truncated $q\bar{q}$ BSE 
\begin{eqnarray}\nonumber
\Gamma(k;P)&=&-\frac{4}{3}\int^\Lambda_q\!\!\!\!\mathcal{G}((k-q)^2)\; D_{\mu\nu}^f(k-q) \;
\gamma_\mu \; \chi(q;P)\;\gamma_\nu \, \,,\\\label{bse}
\chi(q;P)&=&S(q_+) \Gamma(q;P) S(q_-)\,,
\end{eqnarray}
and the rainbow-truncated quark DSE
\begin{eqnarray}\nonumber
S(p)^{-1}  &=&  (i\gamma\escalpr p + m_q)+  \Sigma(p)\,,\\\label{dse}
\Sigma(p)&=& \frac{4}{3}\int^\Lambda_q\!\!\!\! \mathcal{G}((p-q)^2) \; D_{\mu\nu}^f(p-q) 
\;\gamma_\mu \;S(q)\; \gamma_\nu \,. 
\end{eqnarray}
Solution of the BSE yields the
Bethe-Salpeter amplitude (BSA) $\Gamma(k;P)$, which is combined with two dressed quark propagators
into what is commonly referred to as the ``Bethe-Salpeter wave function'' $\chi(q;P)$.
In these equations Dirac and flavor indices have been omitted for simplicity and the factor 
$\frac{4}{3}$ comes from the color trace. $D_{\mu\nu}^f(p-q)$ is the free gluon 
propagator, $\gamma_\nu$ is the bare quark-gluon vertex, $\mathcal{G}((p-q)^2)$ 
is an effective running coupling, and the (anti)quark momenta are $q_+ = q+\eta P$ and $q_- = q- (1-\eta) P$.
$\eta \in [0,1]$ is referred to as the momentum partitioning parameter, which is usually 
set to $1/2$ for systems of equal-mass constitutents.
$\int^\Lambda_q=\int^\Lambda d^4q/(2\pi)^4$ represents a translationally invariant 
regularization of the integral, with the regularization scale $\Lambda$ \cite{Maris:1997tm}.
Furthermore, $\Sigma(p)$ denotes the quark self energy and $m_q$ the current-quark mass.
The solution for the quark propagator $S(p)$ requires a renormalization procedure, the details of which
have been omitted for simplicity but can be found together with the general structure of both the BSE
and quark DSE in \cite{Maris:1997tm,Maris:1999nt}.

The inverse quark propagator has the general form $S(p)^{-1}=i\gamma\cdot p\,A(p^2)+B(p^2)$.
Solving the two coupled integral equations for $A$ and $B$, one obtains an input for 
the $q\bar{q}$ BSE. In particular, in Euclidean momentum space a solution 
of the BSE for a bound-state of mass $M$ implies $P^2=-M^2$, which in turn requires knowledge 
of the quark propagator at the complex momenta $q_\pm$. The corresponding arguments
of $A$ and $B$, $q_\pm^2$, lie inside a parabola defined by the complex points $(-M^2/4,0)$ and 
$(0,\pm M^2/2)$ for $\eta=1/2$ (for a more detailed discussion, see e.\,g.~App.~A of Ref.~\cite{Fischer:2008sp}). 
The analytic continuation required presents a considerable numerical challenge at large quark 
masses (in practice for $m_q$ larger than the charm quark mass). The method used here is based on
a strategy described in \cite{Fischer:2005en}, which is further developed in \cite{Krassnigg:2008gd} with
the help of a particularly useful and stable way to numerically evaluate Cauchy's integral formula \cite{Ioakimidis:1991io}. 

To fully specify the items defined in Eqs.~(\ref{bse}), (\ref{dse}) one needs a form for the 
effective coupling $\mathcal{G}(s)$, $s:=(p-q)^2$. Here the form introduced in Ref.~\cite{Maris:1999nt} is chosen, which is
\begin{equation}\label{coupling}
\frac{{\cal G}(s)}{s} = \frac{4\pi^2 D}{\omega^6} s\;\mathrm{e}^{-s/\omega^2}
+\frac{4\pi\;\gamma_m \pi\;\mathcal{F}(s) }{1/2 \ln [\tau\!+\!(1\!+\!s/\Lambda_\mathrm{QCD}^2)^2]}.
\end{equation}
This Ansatz produces the correct perturbative limit, i.\,e.~it preserves the one-loop 
renormalization group behavior of QCD for solutions of the quark DSE. As given 
in \cite{Maris:1999nt}, ${\cal F}(s)= [1 - \exp(-s/[4 m_t^2])]/s$, $m_t=0.5$~GeV, 
$\tau={\rm e}^2-1$, $N_f=4$, $\Lambda_\mathrm{QCD}^{N_f=4}= 0.234\,{\rm GeV}$, and $\gamma_m=12/(33-2N_f)$.  
\begin{figure}
    \begin{center}
                    \includegraphics[scale=0.35,angle=270,clip=true]{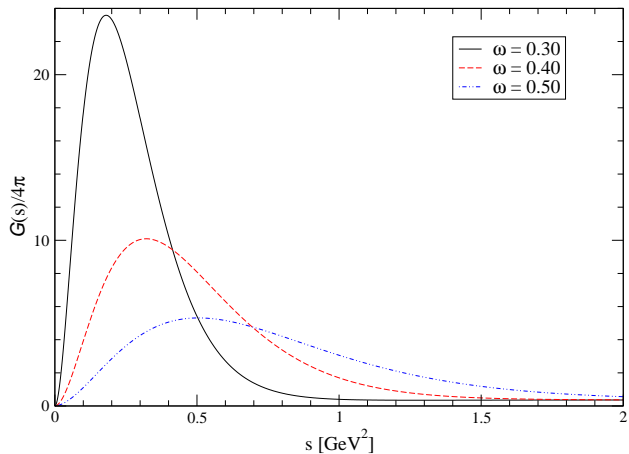}
                    \caption{(Color online) The form of the effective coupling (\ref{coupling})
		    for the three values of the parameter $\omega$ used herein.\label{fig:c}} 
       \end{center}
\end{figure}

$D$ and $\omega$ are in principle free parameters of the model. In \cite{Maris:1999nt} they were fixed 
together with the current-quark mass $m_q$ in Eq.~(\ref{dse}) to the chiral condensate as well as
the pion mass and leptonic decay constant. It emerges from the results presented there that the computed
values for the chiral condensate, $m_\pi$, $f_\pi$, $m_\rho$, and $f_\rho$ change very little, if one 
varies $D$ and $\omega$ such that their product remains constant and $\omega$ lies in the range $[0.3,0.5]$ GeV.
As a result, one can interpret a setup with $D\cdot\omega=const.$ as a one-parameter model, which is 
essentially determined by the requirement to correctly implement chiral symmetry and its dynamical
breaking as well as to obtain the chiral condensate and $f_\pi$ of the correct magnitude. To illustrate
the difference in the coupling generated by the first term in Eq.~(\ref{coupling}), Fig.~\ref{fig:c} shows
the corresponding curves for three values of $\omega$ (and corresponding $D$ for the above interval
boundaries and its center value). It is clear that the main difference between the curves lies in the low- and 
intermediate-momentum range, which corresponds to the long-range part of the interaction.

In the present study the same parameter range is investigated for mesons with all quantum numbers corresponding to 
spin 0 and 1. While in \cite{Maris:1999nt} the current-quark mass was slightly readjusted for each value
of $\omega$ in order to achieve the exact same results for $m_\pi$, this is not done here, since it 
obscures the influence of $D$ and $\omega$ on the spectrum, which is the main point of the investigation.
The parameters are fixed here as follows: the above-mentioned range of  $\omega \in [0.3,0.5]$ GeV is inherited
from \cite{Maris:1999nt} together with the value of $D\cdot\omega=0.372\;\mbox{GeV}^3$ used there. The remaining
parameters, four current quark masses for the flavors $u/d$, $s$, $c$, and $b$, are then each fixed
to the corresponding experimental vector-meson ground-state mass. This last step is motivated by the fact that
for heavy quarkonia, the vector state is the best-known experimentally. For the $\bar{s}s$ case, the vector 
state is again the better choice due to ideal $SU(3)$-flavor mixing in the vector case, since a corresponding
pseudoscalar meson state does not appear experimentally and the RL-truncated BSE kernel does not
contain flavor-mixing processes. To make everything consistent, the light-quark mass is then fixed to the 
$\rho$ meson instead of the pion mass.

\section{Results and Discussion\label{results}}

Equations (\ref{bse}) and (\ref{dse}) are solved consistently in RL truncation using the interaction
of Ref.~\cite{Maris:1999nt}. As laid out in the introduction, this setup has been applied to numerous
observables for pseudoscalar- and vector-meson ground states individually in the literature. 
In the present work, the new aspect is a consistent treatment of all meson states with $J=0,1$ and a 
comprehensive study of the dependence of these states on the parameters $D$ and $\omega$ of the model, 
more precisely with the restriction mentioned above, namely $D\cdot\omega=const.$ As reported earlier
in connection with radially excited states of pseudoscalar mesons in the same setup \cite{Holl:2004fr,Holl:2005vu},
such a study can be used to draw conclusions about the effective range and pointwise form of the
long-range part of the strong interaction between, in particular, also light quarks. 

This is due to the observation that the $\omega$-independence observed for pseudoscalar- and vector-meson
ground-state properties does not survive for radial excitations: for example, meson masses can vary by several
hundred MeV over the range $\omega \in [0.3,0.5]$ GeV (see \cite{Holl:2004fr,Holl:2004un,Krassnigg:2006ps,Krassnigg:2008gd}). 
Basically, it would therefore be possible to use such excited states to fix all parameters in the interaction completely
and attempt a quantitative study and comparison to other approaches/experimental data.
However, this is not the aim here: the present study is almost purely qualitative. It focusses simply on
the effect of the long-range part of the strong interaction, encoded in the parameters of the present model,
on states with various quantum numbers and quark masses. Results have been obtained for
ground states with equal-mass constituents of each $0^{PC} $ and $1^{PC}$ channel, spanning the entire range from 
light to heavy quarks.
\begin{figure}
    \begin{center}
                    \includegraphics[scale=0.35,angle=270,clip=true]{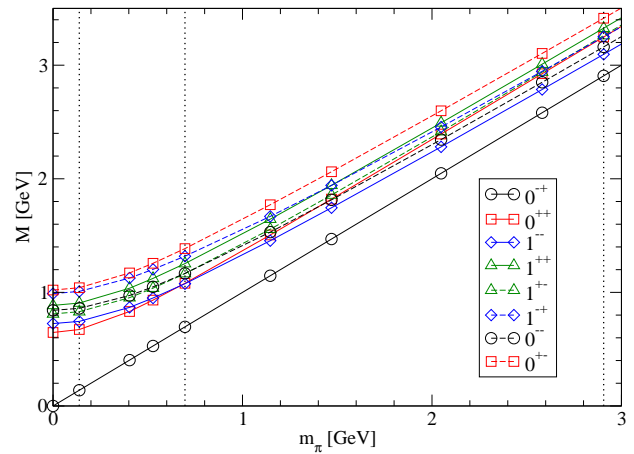}
                    \caption{(Color online) dependence of meson masses on $m_\pi$ (more generally, $m_\pi$ denotes the
		    pseudoscalar-meson mass for a given current-quark mass). Vertical dotted lines
		    correspond to positions for light, strange, and charm $\bar{q}q$ states.\label{fig:1}} 
       \end{center}
\end{figure}

\begin{figure*}
    \begin{center}
                    \includegraphics[scale=0.55,angle=270,clip=true]{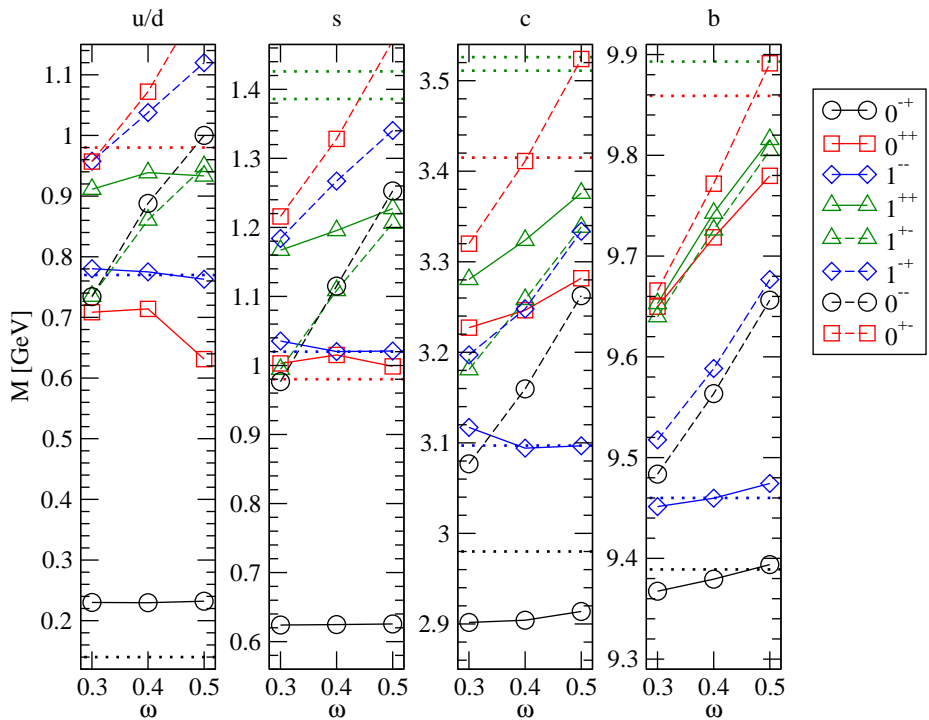}
                    \caption{(Color online) dependence of meson masses on $\omega$. Dotted lines
		    correspond to experimental data \cite{Amsler:2008zzb,Aubert:2008vj}. Note that 
		    there is no flavor mixing in RL truncation, i.\,e.~there is no experimental
		    number for an $\bar{s}s$ pseudoscalar meson.\label{fig:2}} 
       \end{center}
\end{figure*}

To illustrate the evolution of the various meson masses with the quark/pion mass, Fig.~\ref{fig:1} shows
the corresponding results of the calculation from the chiral limit to charmonium. The curves beyond
that value continue to rise linearly as expected. The results for bottomonium are provided in Fig.~\ref{fig:2}.

Two observations from Fig.~\ref{fig:1} are noteworthy. Firstly and most prominently, 
a comment on meson states with ``exotic'' quantum numbers is in order. Meson states with such quantum numbers,
which are not available for a $\bar{q}q$ state in quantum mechanics, appear naturally in a quark-antiquark Bethe-Salpeter
equation. For example, $1^{-+}$ states have previously been studied and discussed using a separable BSE kernel in \cite{Burden:2002ps}
and, for light quark masses, also using the interaction of Ref.~\cite{Maris:1999nt} in \cite{Cloet:2007pi}.
Here they are included for completeness with the immediate remark that corrections to the masses
of these states can be expected to be at least of the same order of magnitude as those for axial-vector mesons, 
whose masses are underestimated by several hundred MeV. Still, for the purpose of Fig.~\ref{fig:2}, their 
inclusion reveals interesting analogies discussed below.

\begin{figure}
    \begin{center}
                    \includegraphics[scale=0.45,angle=270,clip=true]{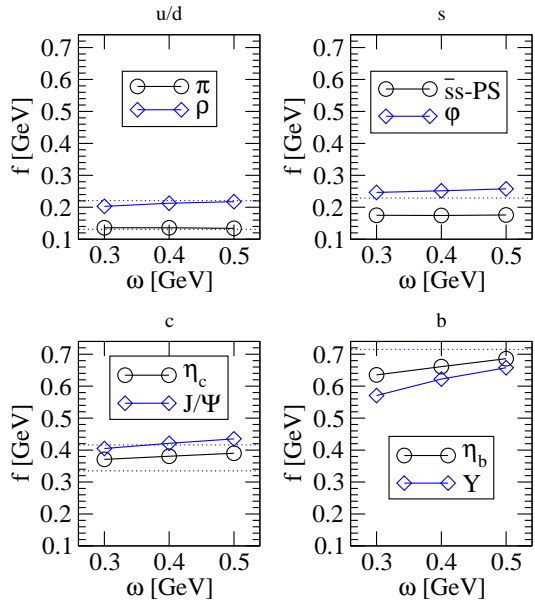}
                    \caption{(Color online) dependence of meson leptonic decay constants on $\omega$. Dotted lines
		    correspond to experimental data \cite{Amsler:2008zzb,Edwards:2000bb}. Note that 
		    there is no flavor mixing in RL truncation, i.\,e.~there is no experimental
		    number for an $\bar{s}s$ pseudoscalar meson decay constant. 
		    Furthermore, measurements for the $\eta_b$ decay constant have not yet been reported.\label{fig:3}} 
       \end{center}
\end{figure}
The other observation from Fig.~\ref{fig:1} to be made here is that the (purely RL-$\bar{q}q$) scalar state
lies below the vector state in the chiral limit; the two switch positions at about the strange-quark mass,
beyond which they confirm with the ordering observed experimentally. Due to the complicated situation for
scalar mesons as well as the simplicity of RL truncation, a meaningful quantitative
comparison of the scalar result at light quark masses with experimental data is beyond the present study.

Figure \ref{fig:2} provides the essential information for the conclusions presented from this study. It
contains all results for meson states with $0^{PC} $ and $1^{PC}$ for the four relevant values of the quark mass
as functions of the model parameter $\omega$.
As has been mentioned above and is apparent from the figure, the vector-meson masses have been fixed to the
corresponding experimental values for the central value of $\omega$. In accordance with the original observation in 
\cite{Maris:1999nt}, pseudoscalar and vector masses depend on $\omega$ only very slightly. It should be stressed again here
that the quark masses were not refitted for different values of $\omega$ in order not to obscure the $\omega$ dependence of
the state. In a quark-model interpretation, with the quark-antiquark orbital angular momentum denoted by
$L$, pseudoscalar and vector states correspond (mainly) to $L=0$ states, whereas
all other states under consideration here have higher $L$ and are therefore orbital excitations; the exotics are constructed
using addtional gluonic degrees of freedom and thus correspond to excitations as well. It is therefore not surprising that
all other states in Fig.~\ref{fig:2} share the characteristic $\omega$ dependence of radial excitations demostrated earlier
\cite{Holl:2004fr,Holl:2004un,Krassnigg:2006ps,Krassnigg:2008gd}. 
An additional aspect of exotic states is helpful in understanding this: they are in fact radial excitations of 
their non-exotic counterparts in a more general setup where the restriction on a particular $C$ parity is lifted.

With this overall picture in mind, it is clear that quantitative studies in this approach must explore the parameter
dependence of the results on their respective interaction. This, in turn, can lead to conclusions about the pointwise
form of that interaction. 

As a further illustration, Fig.~\ref{fig:3} shows the $\omega$ dependence of the pseudoscalar- and vector-meson
leptonic decay constants for all quark masses together with experimental data, 
where known. More precisely, there are no experimental numbers for $f_{\bar{s}s-PS}$ and $f_{\eta_b}$. Note that there 
is a $\pm 75$ MeV error bar on the measurement of $f_{\eta_c}$ \cite{Edwards:2000bb}, which is not
indicated in the figure. Observations from this figure are similar to those from Fig.~\ref{fig:2}, namely that the 
$\omega$ dependence of pseudoscalar- and vector- meson properties in this setup is small. An interesting
detail is that the calculated results show $f_{\eta_b}>f_{Y}$ while for all lower states the vector decay constant is 
larger than the pseudoscalar. However, since the disagreement with the experimental number for $f_Y$ is 
rather large, also this result should be interpreted cautiously and in the context of the present qualitative study.

\section{Conclusions and Outlook\label{conclusions}}
This paper presents a qualitative overview of meson spectra and decay constants for spins $J=0,1$ for 
a sophisticated model of QCD using a rainbow truncation of the quark DSE and the ladder
quark-antiquark BSE. Beyond various studies of aspects of meson properties in this setup existent in the
literature, the results presented here are comprehensive over the whole range of quark masses from the
chiral limit up to the bottom quark and for all quantum numbers possible for $J=0,1$ including so-called
``exotic'' ones, which appear naturally in a Bethe-Salpeter treatment of the quark-antiquark system.
Furthermore, the results' dependence on the model parameters is explored systematically.

The results show a pattern where states corresponding mainly to $L=0$ in the quark model show only a
slight sensitivity to the long-range part of the strong interaction, whereas those corresponding to
$L=1$ or with exotic quantum numbers show a clear dependence, which is analogous to that discovered
previously for radial meson excitations. As a consequence, all excitations can be related to the
long-range details or form of the interaction under consideration, which in principle can allow a 
pointwise investigation of that interaction and makes these states a prime object for further studies. 

Here only states with equal-mass constituents were studied. Subsequent investigations in this approach 
will, among other things, deal with higher spin states as well as states with unequal-mass constituents; 
in particular heavy-light systems are of interest, since they combine the two regimes of heavy and 
light quarks in a unique fashion, thus offering an opportunity to also study in the same detail the 
interaction between quarks of light and heavy masses.

\appendix*

\section{Structure of the BSA}\label{bsa}
The BSA $\Gamma^{(\mu)}$ of a meson as a bound state of a quark-antiquark pair 
depends on their total momentum $P$ as well as the relative momentum $q$. The appearance
of the Lorentz index $\mu$ is related to the spin $J$ of the meson: for $J=0$, the amplitude
has no index, for $J=1$ there is $\mu$. In terms 
of Lorentz-invariant variables, the amplitudes depend on $P^2$, $q^2$, and $q\escalpr P$.
The spin structure is taken into account by the fact that $\Gamma$ is a $4\times 4$ matrix
in spinor space \cite{Smith:1969az}. The corresponding basis can be built from products of Dirac-$\gamma$ matrices. 
The general dependence of the BSA on these variables can be written in terms of 
$N$ covariant structures $T_i$ and scalar amplitudes $F_i$; $i=1,\ldots, N$: 
\begin{eqnarray}\nonumber
\Gamma^{(\mu)} (P;q)=\sum^N_{i=1}T_i^{(\mu)}(P;q;\gamma) F_i(q^2,q\escalpr P,P^2)\;.
\end{eqnarray}
Note that for a bound-state amplitude --- the solution of a homogeneous BSE --- the the on-shell
condition requires $P^2=-M^2$ to be fixed. However, 
one varies $P^2$ in the solution process of the homogeneous BSE to find the above condition.
In the corresponding inhomogeneous BSE one has $P$ and therefore also $P^2$ as a completely 
independent variable.

For $J=0$, i.e.~scalar or pseudoscalar mesons, the BSA contains four independent covariant structures.
For a scalar, on can choose
\begin{equation}\label{covsc}
\begin{array}{ll}
T_1 = \mathbf{1} & T_2 = i\,\gamma\escalpr P \\
T_3 = i\,\gamma\escalpr q & T_4 = i\,\sigma^{q,P} 
\end{array}
\end{equation}
where $\sigma^{q,P}:=i/2\,[\gamma\cdot q,\gamma\cdot P]$.
Note that these covariants are in general neither orthogonal nor normalized in terms
of the two four-momenta $q$ and $P$ (for example, one could have $T_3= \frac{i}{\sqrt{q^2}}\,\gamma\escalpr q$).
The scalar product for general covariants is defined via the Dirac trace
\begin{equation}
\sum_{(\mu)} \mathrm{Tr} [T_i^{(\mu)} T_j^{(\mu)}]=t_{ij}f(i,j)\;,
\end{equation}
where for an orthogonal basis one has $t_{ij}=\delta_{ij}$ and $f(i,j)$ are functions
of $q^2$, $P^2$, and $q\escalpr P$, and the sum over $\mu$ is carried out only for $J=1$.

The scalar amplitudes $F_i(q^2,q\escalpr P,P^2)$ depend on two independent variables, 
since the total momentum-squared $P^2$ is fixed for an on-shell meson. The relative-momentum
squared $q^2$ corresponds to a radial variable in Euclidean momentum space, while the
scalar product $q\escalpr P$ can be understood in terms of the cosine $z$ of an angle between
$q$ and $P$ by writing $q\escalpr P=\sqrt{q^2\,P^2}z$. The amplitudes $F_i$ can be decomposed
by a Chebyshev expansion as
\begin{equation}\label{cheby}
F_i(q^2,q\escalpr P,P^2)=\sum_{j=0}^\infty\;{}^jF_i(q^2,P^2)\;U_j(z)\;,
\end{equation}
where $U_j(z)$ are the Chebyshev polynomials of the second kind and the ${}^jF_i$ are the
corresponding Chebyshev moments, which effectively only depend on $q^2$. An illustration
of ground- and excited-state pseudoscalar meson Chebyshev moments can be found in \cite{Krassnigg:2003dr}.

A scalar meson with equal-mass constituents has the 
quantum numbers $J^{PC}=0^{++}$. The BSA as defined above with the covariants for the
scalar case has $J^P=0^+$, but is not restricted \emph{a priori} to either $C=+1$ or $C=-1$.
When using a momentum partitioning parameter of $\eta=1/2$, one can obtain an amplitude 
with positive $C$-parity by restricting the $C$-parity for each $F_i$ according to the
value of $C$ for the corresponding $T_i$ (see also \cite{Maris:1997tm,Maris:1999nt}).
For $T_i$ odd under $C$ one has to ensure that $F_i$ is odd as well. This is done via 
the dependence on $q\escalpr P$, which is odd under $C$. In the Chebyshev expansion defined in 
Eq.~(\ref{cheby}) this means that for an even/odd $F$ one would keep only even/odd terms in the expansion.
In this way, it is immediately clear how to construct an amplitude with the opposite, in some
cases called ``unnatural'' charge-parity, namely by making all odd $F$s even and vice versa.
In this way, the BSE can be used to study mesons with ``exotic'' quantum numbers (see also 
\cite{Burden:2002ps,Jarecke:2002xd}).

For the scalar basis, $T_2$ is odd under $C$, the others are even. As a result, $F_2$ is
an odd function of $q\escalpr P$, the others are even. With the above choice of covariants 
(and as mentioned above $\eta=1/2$),
the Chebyshev moments of $F_2$ are purely imaginary, the others are real. One
could unify all covariants' behavior by making the modification
\begin{equation}
T_2 = i\,\gamma\escalpr P \rightarrow i\,q\escalpr P\;\gamma\escalpr P\;,
\end{equation}
after which all covariants are even and all amplitudes real.
 It should be noted
here that in a more general setup, which is e.g.~needed in the case of unequal-mass constituents,
a restriction like the above is not possible, since such a state is not a $C$-parity eigenstate.
In this case, both odd and even Chebyshev moments will contribute in general, i.e.~the $F_i$ will be
complex. However, in the limiting case of equal constituent masses, the real/imaginary pattern described
above is recovered also numerically.

The basis used for a pseudoscalar $J^P=0^-$ meson is
\begin{equation}\label{covps}
\begin{array}{ll}
T_1 = i\,\gamma_5 & T_2 = \gamma_5\;\gamma\escalpr P \\
T_3 = \gamma_5\;\gamma\escalpr q & T_4 = \gamma_5\;\sigma^{q,P} 
\end{array}
\end{equation}
where $T_3$ is odd under $C$ and the others are even. As a result, for a state with
$J^{PC}=0^{-+}$, $F_3$ must be odd and the others even, and the opposite for the exotic $J^{PC}=0^{--}$.

For $J=1$ one has 12 independent covariant structures in the BSA. Since an on-shell (axial-)vector meson is
transverse (i.e.~$P_\mu\Gamma^\mu(P;q)\stackrel{!}{=}0$), 8 (transverse) covariants remain. With the definitons of the 
transversely projected
\begin{eqnarray}
{q^T}^\mu&:=&q^\mu-\frac{q\escalpr P}{P^2}P^\mu\\
{\gamma^T}^\mu&:=&\gamma^\mu-\frac{\gamma\escalpr P}{P^2}P^\mu
\end{eqnarray}
one arrives at a simple basis for a $J^P=1^-$ state by choosing
\begin{equation}\label{covve}
\begin{array}{ll}
T_1^\mu = {\gamma^T}^\mu & T_2^\mu = {q^T}^\mu\,\gamma\escalpr q \\
T_3^\mu = {q^T}^\mu\,\gamma\escalpr P & T_4^\mu = -i\,{\gamma^T}^\mu\,\sigma^{q,P}-{q^T}^\mu\,\gamma\escalpr P \\
T_5^\mu = i\,{q^T}^\mu\,\mathbf{1} & T_6^\mu = i\,({\gamma^T}^\mu\,\gamma\escalpr q-\gamma\escalpr q \,{\gamma^T}^\mu)\\
T_7^\mu = i\,{\gamma^T}^\mu\,\gamma\escalpr P & T_8^\mu = {q^T}^\mu\,\sigma^{q,P} 
\end{array}
\end{equation}
$T_3$ and $T_6$ are even under $C$, the others are odd. To obtain a $J^{PC}=1^{--}$ state, 
$F_3$ and $F_6$ must be odd functions of $q\escalpr P$, the others even. For the exotic 
$J^{PC}=1^{-+}$ state, the situation is reversed.

For the axialvector $J^P=1^+$ meson, the basis used can be easily constructed from the 
vector case above by multiplication with $\gamma_5$ in analogy to the $J=0$ case. One has
\begin{equation}\label{covav}
\begin{array}{ll}
T_1^\mu = i\gamma_5\;{\gamma^T}^\mu & T_2^\mu = i\gamma_5\;{q^T}^\mu\,\gamma\escalpr q \\
T_3^\mu = i\gamma_5\;{q^T}^\mu\,\gamma\escalpr P & T_4^\mu = \,\gamma_5\;{\gamma^T}^\mu\,\sigma^{q,P}-i\gamma_5\;{q^T}^\mu\,\gamma\escalpr P \\
T_5^\mu = \,\gamma_5\;{q^T}^\mu & T_6^\mu = \gamma_5\;({\gamma^T}^\mu\,\gamma\escalpr q- \gamma\escalpr q\,{\gamma^T}^\mu)\\
T_7^\mu = \,\gamma_5\;{\gamma^T}^\mu\,\gamma\escalpr P & T_8^\mu = -i\gamma_5\;{q^T}^\mu\,\sigma^{q,P} 
\end{array}
\end{equation}
For the axialvector case, both sets $J^{PC}=1^{+-}$ and $J^{PC}=1^{++}$ are ``non-exotic'', i.\,e.~they
can be constructed in a constituent-quark model as a pure quark-antiquark state.
$T_3$, $T_5$, $T_7$, and $T_8$ are odd under $C$; to obtain $J^{PC}=1^{++}$ one needs
the corresponding amplitudes to be odd functions of $q\escalpr P$, the others even. Again,
for $J^{PC}=1^{+-}$ the situation is reversed.

\begin{acknowledgments}
The author would like to acknowledge valuable discussions with M.~Blank, G.~Eichmann, 
C.\,D.~Roberts, and M.~Schwinzerl.
This work was supported by the Austrian Science Fund \emph{FWF} under project no.\ P20496-N16 and
benefited from the computing resources of the Argonne National Laboratory LCRC.
\end{acknowledgments}


\end{document}